\documentclass[prb,preprint,draft,amsmath,showpacs]{revtex4}
\usepackage{graphics}
\begin{document}
%%%%%%%%%%%%%%%%%%%%%%%%%%%%%%%%%%%%%%%%%%

\bibliographystyle{prsty}
\input epsf

\title {Optical evidence for a spin-filter effect in the charge 
transport 
of $Eu_{0.6}Ca_{0.4}B_{6}$}

\author {A. Perucchi, G. Caimi and H.R. Ott}
\affiliation{Laboratorium f\"ur Festk\"orperphysik, ETH Z\"urich,
CH-8093 Z\"urich, Switzerland}\

\author {L. Degiorgi}
\affiliation{Paul Scherrer Institute, CH-5232 Villigen and 
Laboratorium f\"ur Festk\"orperphysik, ETH Z\"urich,
CH-8093 Z\"urich, Switzerland}\

\author {A.D. Bianchi and Z. Fisk}
\affiliation{NHMFL-FSU, Tallahassee FL 32306, U.S.A.}

\date{\today}

\begin{abstract}
We have measured the optical reflectivity $R(\omega)$ of $Eu_{0.6}Ca_{0.4}B_{6}$ as a function of temperature 
between 1.5 and 300 $K$ and in external magnetic fields up to 7 $T$. The 
slope at the onset of the plasma edge 
feature in $R(\omega)$ increases with decreasing 
temperature and increasing field but the plasma edge itself does not 
exhibit the remarkable blue shift that is observed in the 
binary compound $EuB_{6}$. The analysis of the magnetic field 
dependence of the low temperature optical conductivity spectrum confirms the 
previously observed exponential decrease of the electrical resistivity 
upon increasing, field-induced bulk magnetization at constant 
temperature. In addition, the individual exponential magnetization dependences of 
the plasma frequency and scattering rate are extracted from the 
optical data.  
\end{abstract}
\maketitle

%\section{Introduction}
Remarkable variations of electronic transport properties, distinctly 
depending on the bulk magnetization, have been observed for 
materials, such as rare-earth hexaborides and manganites \cite{ref1,ref2}. A very direct link between electronic transport and bulk 
magnetization $M$ was revealed by measurements of magneto-optical 
properties of ferromagnetic $EuB_{6}$ (Refs. \onlinecite{ref3} and 
\onlinecite{ref4}), exhibiting a substantial blue shift of the plasma edge in the 
optical reflectivity with decreasing temperature and increasing 
magnetic field $H$ (Refs. \onlinecite{ref5} and \onlinecite{ref6}).

More recently, dc magneto-transport and magnetization experiments on 
a series of $Eu_{1-x}Ca_{x}B_{6}$ compounds provided results that 
again reflect the intimate relation between the electronic 
conductivity and the magnetization. In particular, for material with 
x=0.4, an exponential decrease of the resistivity $\rho_{dc}$ as a function of magnetization at constant 
temperature close to and below the Curie temperature $T_{C}$ was reported \cite{ref7}. 
Based on these results it was suggested that some kind of 
magnetization-dependent barriers or intrinsic spin-filter 
effects dominate the electronic transport in the magnetically ordered phase of 
this material \cite{ref7}. 

%<<<<<<<<<<<<<<<<<<<<<<<< FIGURE 1 >>>>>>>>>>>>>>>>>>>>>>>>>
\begin{figure}[t]
  \begin{center}
   \leavevmode
   \epsfxsize=8cm \epsfbox {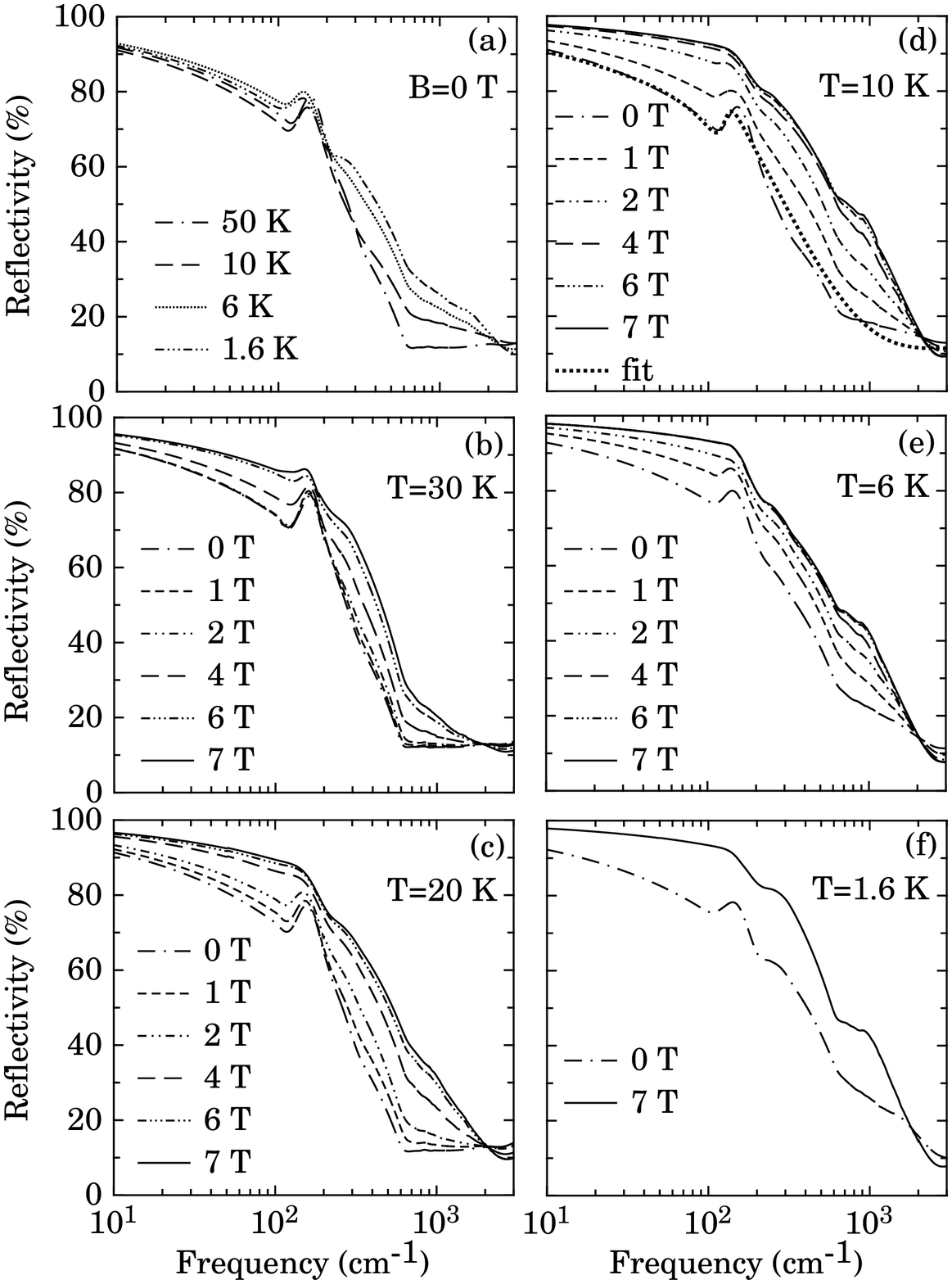}
    \caption{(a) Optical reflectivity $R(\omega)$ of $Eu_{0.6}Ca_{0.4}B_{6}$ in zero field at 
temperatures between 50 and 1.6 $K$. (b)-(f) Magnetic field dependence of $R(\omega)$ at 
selected temperatures. In panel (d) the Lorentz-Drude 
phenomenological fit at 10 $K$ and 0 $T$ is also reproduced in order 
to demonstrate the fit quality.}
\label{MORefl}
\end{center}
\end{figure}
%<<<<<<<<<<<<<<<<<<<<<<<< figure 1 >>>>>>>>>>>>>>>>>>>>>>>>>

It therefore seemed of interest to further characterize the charge transport 
in $Ca$-doped $EuB_{6}$ by investigating the electrodynamic response. 
Since $\rho_{dc}\sim \Gamma/\omega_{p}^{2}$, where $\Gamma$ and 
$\omega_{p}$ are the scattering rate and the plasma frequency of 
the itinerant charge carriers, respectively, it is instructive to 
single out the relevance of those quantities in relation with the dc 
properties. The measured optical response and its phenomenological analysis, based on the classical dispersion theory, 
offers this possibility and provides information on important electronic 
parameters, including the disorder-induced scattering rate of 
the charge carriers.

The single-crystal of $Eu_{0.6}Ca_{0.4}B_{6}$ was prepared by 
solution growth from $Al$ flux, using the necessary high purity elements as starting 
materials. Our specimen (1.7x1.7x0.5 $mm^{3}$) is from the same thoroughly 
characterized batch of samples described in Ref. 
\onlinecite{ref7}. From the absence of any sharp features in $\rho_{dc}(T)$ at 1.2 
$K$, the critical temperature for superconductivity of $Al$, we 
exclude the presence of $Al$ inclusions in the sample \cite{ref7}. The 
optical reflectivity $R(\omega)$ was measured in a broad spectral range from 
the far infrared to the ultraviolet and as a function of both temperature and 
magnetic field. The corresponding experimental details are described 
in Refs. ~\onlinecite{ref5} and \onlinecite{ref6}.

Figure 1 summarizes the relevant reflectivity $R(\omega)$ results, at selected 
temperatures from above to below $T_{C}$ and as a function of the external 
magnetic field. We limit our presentation 
to the spectral range from the far infrared (FIR) up to the mid-infrared, where 
the field and temperature induced variations of $R(\omega)$ are most 
prominent. Around 5000 ~$cm^{-1}$ 
all recorded $R(\omega)$ spectra merge and above 50 $K$ no field 
dependence is observed. A comparison with previously 
published data for $EuB_{6}$ (Ref. \onlinecite{ref6}) 
reveals distinct differences. Although the reflectivity is still of 
metallic type, the onset of its plasma edge in zero field is broad in 
$Ca$-doped $EuB_{6}$, i.e., a much less sharp feature than the increase 
of $R(\omega)$ with decreasing $\omega$ for $EuB_{6}$. Another difference is 
the much less pronounced blue shift of the plasma edge with decreasing temperature or 
increasing field in the $Ca$-doped material in comparison with the 
features of the binary compound \cite{ref6}. The peak at about 150 
$cm^{-1}$ is an optically active phonon \cite{ref5}. A closer 
inspection of the data reveals a small shift of this phonon mode to lower 
frequencies at temperatures below 6 $K$. The other 
infrared active phonon, observed around 850 $cm^{-1}$ in $EuB_{6}$ (Refs. 
\onlinecite{ref5} and \onlinecite{ref6}), can barely be identified 
here. It appears as a tiny spike, almost completely screened by the 
plasma edge. Since it is of no relevance in this work, we do not discuss 
it further. At fixed temperature but increasing magnetic field, the 
reflectivity is, overall, progressively enhanced, thus screening the phonon mode at 150 $cm^{-1}$. The $R(\omega)$ enhancement 
results mostly from an increasing slope of the plasma edge feature with increasing field. With decreasing temperature and 
increasing field we note the appearance of a pronounced shoulder at about 
1000 $cm^{-1}$, overlapping the plasma edge. We also remark that the behaviour 
of $R(\omega)$ in zero field (Fig. 1a) and decreasing temperature is qualitatively similar to $R(\omega)$ at 
fixed temperature and increasing field, i.e., it steadily increases with 
decreasing temperature. Although not presented here,
similar optical $R(\omega)$ results were obtained for other 
concentrations of $Ca$-doping (e.g., 10$\%$ and 20$\%$).

Below 30 $cm^{-1}$, our lower experimental limit, the Hagen-Rubens (HR) extrapolation \cite{ref8} 
was used for the extension of $R(\omega)$ towards zero frequency. In 
the FIR spectral range the relative 
increase of $R(\omega)$ at fixed temperature but increasing magnetic 
field qualitatively agrees with the trend observed in the 
dc electronic transport data \cite{ref7}, which exhibits a significant negative 
magnetoresistance below 50 $K$. Unexpected and puzzling is the 
temperature dependence of 
$R(\omega)$ in zero field, shown in Fig. 1a. At the lowest 
experimentally accessible frequency, $R(\omega)$ is consistent with the HR extrapolation 
using, however, $\sigma_{dc}$ values that are not in agreement with those of 
Ref. \onlinecite{ref7}. The $\sigma_{dc}$ values used in the HR 
extrapolation increase instead of decreasing by a 
factor of four between 20 and 2 $K$. Inserting the $\sigma_{dc}$ values of Ref. 
\onlinecite{ref7} would imply that $R(\omega)$ at low frequencies and in zero field decreases with decreasing temperature. 
This is not compatible with the experimental findings, at least above 
30 $cm^{-1}$  (Fig. 1a). Optical experiments in the microwave range, 
i.e. below 30 $cm^{-1}$, may eventually solve this puzzle and 
reconcile the optical results with the dc transport properties. Our 
interest here is focussed on the relative magnetic field dependence, which 
does not suffer from the inconsistencies between dc and dynamical 
properties. Therefore, we have not attempted to extrapolate the 
optical data in zero field by forcing an agreement with $\sigma_{dc}$ of 
Ref. \onlinecite{ref7}. As it will be clear below, our choice for 
$\sigma_{dc}$ in zero field neither alters the main content of our 
discussion nor does it affect the conclusion of our work.

%<<<<<<<<<<<<<<<<<<<<<<<< FIGURE 2 >>>>>>>>>>>>>>>>>>>>>>>>>
\begin{figure}[h]
  \begin{center}
   \leavevmode
   \epsfysize=14cm \epsfbox {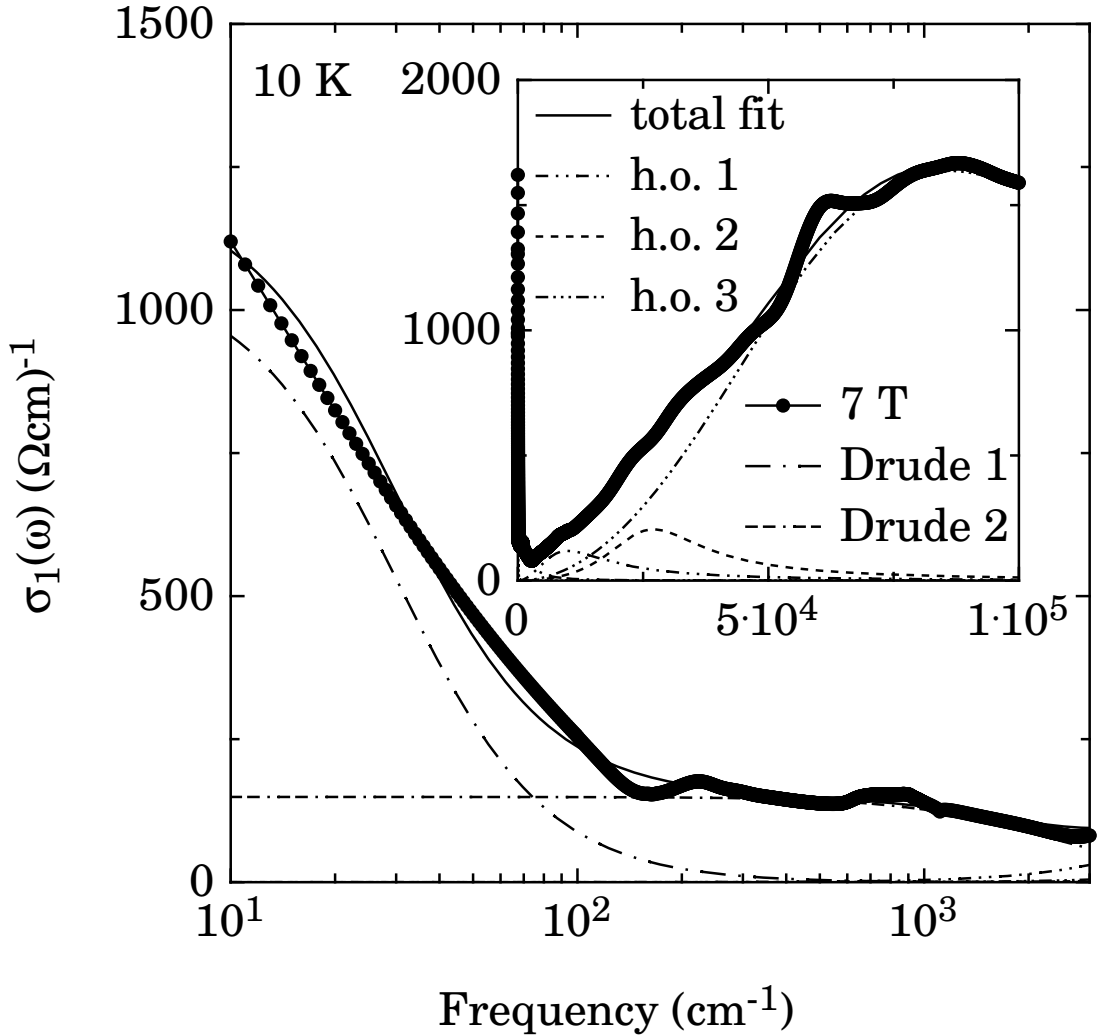}
    \caption{Real part $\sigma_{1}(\omega)$ of the optical conductivity at 10 $K$ and 7 
$T$ with the fit components. The main panel emphasizes the spectral 
range of the two Drude terms, while the inset displays the spectral 
range pertinent for the three harmonic 
oscillators (h.o.). The component of the phonon mode is not shown here. 
The parameters are ($cm^{-1}$): (Drude plasma frequencies) 
$\omega_{p}^{D1}=1382$, $\omega_{p}^{D2}$=4633; (Drude scattering 
rates) $\Gamma_{1}=30$, $\Gamma_{2}=2401$; (h.o. 
strengths) $\omega_{p1}=11207$, $\omega_{p2}=17070$, $\omega_{p3}=111627$; (h.o. scattering rates)  
$\Gamma_{01}=17635$, $\Gamma_{02}=23717$, $\Gamma_{03}=126874$; (h.o. resonance 
frequencies) $\omega_{01}=9986$, $\omega_{02}=27058$, $\omega_{03}=85844$ (Ref. \onlinecite{param}).}
\label{sigma1/fit}
\end{center}
\end{figure}
%<<<<<<<<<<<<<<<<<<<<<<<< figure 2 >>>>>>>>>>>>>>>>>>>>>>>>>

%<<<<<<<<<<<<<<<<<<<<<<<< FIGURE 3 >>>>>>>>>>>>>>>>>>>>>>>>>
\begin{figure}[h]
  \begin{center}
   \leavevmode
   \epsfxsize=6cm \epsfbox {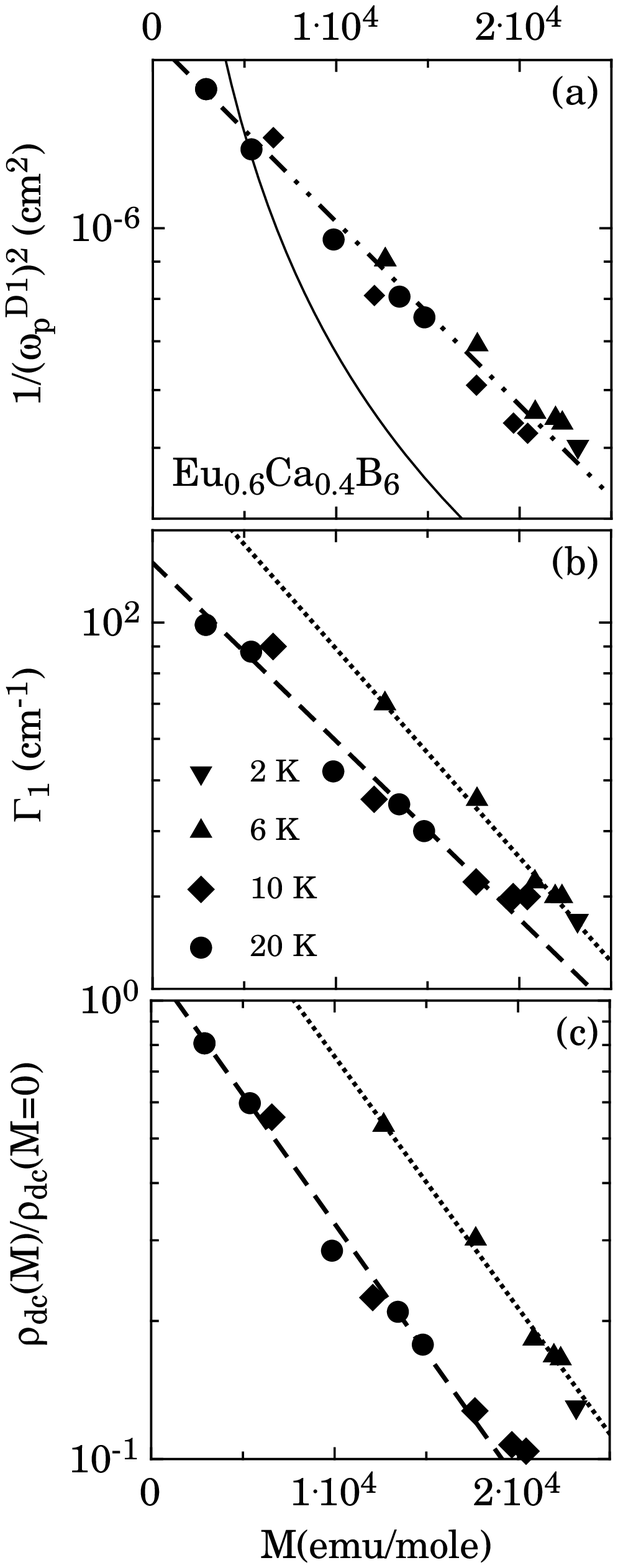}
    \caption{Inverse squared plasma frequency $1/(\omega_{p}^{D1})^{2}$ 
    of the first Drude term as 
a function of the magnetization $M$. The dash-dot-dot line represents eq. 
(1) with the parameters $\omega_{p0}$=744 $cm^{-1}$ and 
$\beta_{\omega}$=5.75x10$^{-5} ~(emu/mole)^{-1}$. The inadequate fit with 
$(\omega_{p}^{D1}(M))^{2} \sim M$ (as for $EuB_{6}$, Ref. 
\onlinecite{ref6}) is also shown as thin solid line.  (b) Scattering rate $\Gamma_{1}$ of 
the first Drude term as a function of $M$. The dotted and the dashed lines 
represent eq. (2) with the parameters $\Gamma_{0}=224 cm^{-1}$ 
and $\beta_{\Gamma}$=9.16x10$^{-5} ~(emu/mole)^{-1}$ for 2 and 6 $K$, and 
$\Gamma_{0}$=130 $cm^{-1}$ and 
$\beta_{\Gamma}$=7.82x10$^{-5} ~(emu/mole)^{-1}$ for 10 and 20 $K$, 
respectively. (c) $\rho_{dc}(M)/\rho_{dc}(M=0)$ versus $M$, 
calculated from the Lorentz-Drude fit parameters as described in the 
text. The dashed and the dotted lines are calculations based on eq. (3) with parameters 
$A=1.19$ and $\beta$=1.30x10$^{-4} ~(emu/mole)^{-1}$ for 10 and 20 $K$, 
and $A=2.67$ and $\beta$=1.26x10$^{-4} ~(emu/mole)^{-1}$ for 2 and 6 $K$, 
respectively. It is obvious that $\beta\sim \beta_{\Gamma}+\beta_{\omega}$ 
(see text).}
\label{exp}
\end{center}
\end{figure}
%<<<<<<<<<<<<<<<<<<<<<<<< figure 3 >>>>>>>>>>>>>>>>>>>>>>>>>

The Kramers-Kronig transformations \cite{ref8} of $R(\omega)$ allow 
us to extract the optical functions, such as the real part 
$\sigma_{1}(\omega)$ of the optical conductivity. As an example we show $\sigma_{1}(\omega)$ at 
10 $K$ and 7 $T$ in Fig. 2. At low frequencies one can easily 
recognize the metallic component of $\sigma_{1}(\omega)$, where the 
limit $\sigma_{1}(\omega \to 0)$ reflects the dc conductivity 
\cite{refcom}. Above 100 
$cm^{-1}$ and up to the mid-infrared, $\sigma_{1}(\omega)$ saturates to an almost constant value. Above $10^{4} 
~cm^{-1}$ (inset of Fig. 2) we note the clear onset of 
interband transitions \cite{ref5,ref6}. The $R(\omega)$ and 
$\sigma_{1}(\omega)$ spectra, for any combination of temperature and 
magnetic field, can be well reproduced by employing the Lorentz-Drude model based on the classical dispersion theory \cite{ref6,ref8}. 
Figure 2 also displays the different components resulting from the fit. Apart from two temperature and field independent Lorentz harmonic 
oscillators (h.o. 2 and 3) in the frequency
range above 1.6x10$^{4}$ ~$cm^{-1}$ (2 $eV$), we consider a temperature and field dependent 
h.o. 1 at about 9900 ~$cm^{-1}$ (1.2 $eV$). Also the phonon mode at 150 $cm^{-1}$ was
described with a h.o. In order to account for 
the optical properties in the far and mid-infrared spectral range 
(i.e., $\omega <$ 3000 $cm^{-1}$), two Drude resonances must be considered. 
For simplicity, we will call them first (D1) and second (D2) Drude 
term, respectively. The total spectral weight encountered in the 
metallic component of $\sigma_{1}(\omega)$, i.e., the sum of the 
squared plasma frequencies of the two Drude terms, defines the 
effective plasma frequency of $Eu_{0.6}Ca_{0.4}B_{6}$. The two-Drude terms suggest a scenario where two different types of charge 
carriers exist. Those charge carriers belong to two bands (e.g., a conduction and an impurity band), 
are differently affected by the ferromagnetic transition, i.e., 
are differently coupled to the spin system (see below), 
and are characterized by distinct scattering rates and effective masses. 

We fit the optical properties by allowing for changes of the plasma frequency of both Drude terms 
($\omega_{p}^{D1}$ and $\omega_{p}^{D2}$, respectively), of the strength 
of the h.o. 1 ($\omega_{p1}$) and the scattering rate ($\Gamma_{1}$) of 
the first Drude term \cite{param}. The first Drude term is characterized by a rather small,
temperature and field-dependent scattering rate ($\Gamma_{1}$). The second 
Drude component mainly accounts for the spectral weight background of 
$\sigma_{1}(\omega)$ causing the shoulder 
most prominently seen at 1000 $cm^{-1}$ in $R(\omega)$ at low temperatures and 
high fields (Fig. 1). Its scattering rate ($\Gamma_{2}$) is large and does not 
vary with temperature and field. Upon decreasing temperature or 
increasing field, both Drude terms acquire spectral weight, which is transferred 
down from higher energies, i.e., from the h.o. 1. As a consequence of the increasing screening by the 
itinerant carriers, the phonon mode component disappears with increasing 
field or decreasing temperature.

Figure 3 captures the main results of this study. In panel (a) we plot $1/(\omega_{p}^{D1})^{2}$
as a function of $M$ (Ref. ~\onlinecite{ref9}), while panel 
(b) shows the scattering rate $\Gamma_{1}(M)$ of the first 
Drude term (Fig. 2). At all temperatures below 30 $K$, we find that these two 
quantities may be well described by
\begin{equation}
    (\omega_{p}^{D1}(M))^{-2}=(\omega_{p0})^{-2}e^{-\beta_{\omega}M}
\end{equation}
and
\begin{equation}
    \Gamma_{1}(M)=\Gamma_{0}e^{-\beta_{\Gamma}M},
\end{equation}
where $\omega_{p0}$ and $\Gamma_{0}$ as well as $\beta_{\omega}$ or 
$\beta_{\Gamma}$ are constants whose values are listed in 
the caption of Fig. 3. From the Lorentz-Drude fit parameters we can calculate 
the conductivity at $\omega$=0 from 
$\sigma_{1}(\omega=0)=\sigma_{1}^{Drude1}(\omega=0)+\sigma_{1}^{Drude2}(\omega=0)=\sigma_{dc}$, and 
$\rho_{dc}=1/\sigma_{dc}$. Panel (c) shows the $\rho_{dc}$ values, normalized 
by $\rho_{dc}(M=0)$ at each temperature, as a 
function of $M$. On purpose we have chosen the same representation of 
$\rho_{dc}(M)/\rho_{dc}(M=0)$ versus $M$ as first suggested 
empirically in Ref. \onlinecite{ref7}. The functional 
form of eq. (1) is the best possible choice to fit the Drude weight 
variation. The choice of $(\omega_{p}^{D1}(M))^{2}$ varying linearly 
with $M$ (as for $EuB_{6}$, Ref. \onlinecite{ref6}) is obviously inadequate.  

It may immediately be seen from Fig. 3c that the expression
\begin{equation}
    \rho_{dc}(M)/\rho_{dc}(M=0)=Ae^{-\beta M}
\end{equation}
describes very nicely the $\rho_{dc}$ values calculated from the 
Lorentz-Drude fit parameters at $\omega=0$. It turns out 
that the best fit to the data points in panel (c) employing eq. (3), 
is very close to the product of eq. (1) and (2), such that 
$\beta\sim\beta_{\omega}+\beta_{\Gamma}$. Both $\beta_{\Gamma}$ and 
$\beta_{\omega}$ are of equal magnitude, suggesting that eq. (1) and (2) 
are of equal importance in shaping the exponential behaviour 
of eq. (3). The difference in $\rho_{dc}(M)/\rho_{dc}(M=0)$ for 
temperatures below 6 $K$ and well above $T_{C}$, is mainly 
due to the differences of the corresponding scattering rates 
($\Gamma_{1}$) (Fig. 3b). Therefore, 
the first Drude term is the most relevant component in determining the dc 
($\omega\to 0$) transport properties and governs to a 
great extent the temperature and magnetic field dependence of 
$\sigma_{1}(\omega)$ at low frequencies. 
With respect to the dc properties, the contribution from the second Drude term is only of marginal 
relevance.  

The two temperature regimes identified in Fig. 3c for temperatures 
below 6 $K$ and well above $T_{C}$ have 
been found in Ref. \onlinecite{ref7} as well, even though $\rho_{dc}(M)/\rho_{dc}(M=0)$ 
was larger at 12 $K$ than at 2 $K$ (see Fig. 4 in Ref. 
\onlinecite{ref7}). This is the consequence of using the 
$\rho_{dc}(M=0)$ values for the normalized representation in Fig. 
3c as they follow from the optical results. These, as stated above, do 
not track the measured dc transport data \cite{ref7}.

Wigger {\it et~al.} suggested that some intrinsic 
magnetization dependent spin-filter effect dictates the electronic 
conduction in this material. The exponential variation of 
$\rho_{dc}(M)$ (eq. (3)) was viewed as being due to magnetization dependent 
tunnel barriers, caused by the domain 
walls of the ferromagnetically ordered phase in zero external field, 
which limit the motion of 
electrons. The domain walls are thought to be favored by the disorder 
on the cation-sublattice and to be progressively weakened and removed 
upon increasing magnetization \cite{ref7}. 
This disorder is most likely also responsible for the rather broad onset 
of the plasma edge feature in $R(\omega)$, at least at high 
temperatures and low fields (Fig. 1). 
The results and the conclusions which follow from the analysis of the optical 
data (Fig. 3) confirm the results obtained in the previous 
investigation of dc magnetotransport 
properties \cite{ref7}. In addition, however, they allow for an analysis of the 
individual magnetization dependences of the two quantities that 
determine the dc transport, i.e., the plasma 
frequency and the scattering rate. The exponential variations that 
link the magnetization to the Drude spectral weight, which is proportional 
to the ratio $n/m$ of the charge carrier concentration ($n$) and the
effective mass ($m$), and to the scattering rate, is rather intriguing. It remains to be seen how these 
findings can be explained on a microscopic level.

\acknowledgments
The authors wish to thank J. Mueller for technical help and G.A. 
Wigger and R. Monnier for fruitful discussions. This work 
has been supported by the Swiss National Foundation for the 
Scientific Research.

\newpage


\begin{thebibliography}{10}
\bibitem{refcom}
Although not shown explicitly, we note that the low frequency part ($\omega<100 ~cm^{-1}$) of $\sigma_{1}(\omega)$ 
narrows with decreasing temperature and increasing field. This is the 
consequence of the enhanced $R(\omega)$ at low frequencies with 
increasing field and decreasing temperature (Fig. 1).
\bibitem{ref9}
The magnetization has been measured up to 5.5 $T$ (Ref. ~\onlinecite{ref7}). 
By assuming a Brillouin function for the overall behaviour of 
$M(H,T)$, a simple fit of the measured magnetization data leads to the 
values of $M$ at 6 and 7 $T$ for different temperatures (We specially 
thank G.A. Wigger for performing these fits). These extrapolated 
values were then combined with the corresponding optical 
parameters ($\omega_{p}$ and $\Gamma$), obtained from the optical 
spectra up to 7 $T$.
\end{thebibliography}
\end{document}